\documentclass[fleqn,twoside]{article}
\usepackage[headings]{espcrc2}

\readRCS
$Id: espcrc2.tex,v 1.2 2004/02/24 11:22:11 spepping Exp $
\usepackage{graphicx}
\usepackage[figuresright]{rotating}

\newcommand{\AmS}{{\protect\the\textfont2
  A\kern-.1667em\lower.5ex\hbox{M}\kern-.125emS}}

\title{Experimental results on $K_L$ decays}

\author{W. Wi\'slicki (on behalf of NA48 at CERN) \address[MCSD]{A. So\l tan Institute for Nuclear Studies,
        Ho\.za 69, PL-00-681 Warszawa} }
       
\runtitle{$K_L$ decays}
\runauthor{W. Wi\'slicki}

\begin{document}

\begin{abstract}
Recent measurements by the NA48 at CERN of selected decays of $K_L$ are presented.
These are the branching ratios and form factors for channels $K_L\rightarrow \pi^{\pm}\pi^0 e^{\mp}\nu_e$ (final result), $K_L\rightarrow\pi^{\pm}e^{\mp}\nu_e$ and $K_L\rightarrow e^+e^-e^+e^-$ (preliminary results).
\vspace{1pc}
\end{abstract}

\maketitle

\section{INTRODUCTION}

The NA48 experiment at CERN, originally designed for the measurement of the direct CP violation \cite{na48_cp}, performs also a rare-decay programme for the $K_S$ and $K_L$ mesons.
Here we report on recent measurements of the decays $K_L\rightarrow \pi^{\pm}\pi^0 e^{\mp}\nu_e$ ($K_{e4}$) \cite{na48_ke4}, $K_L\rightarrow\pi^{\pm}e^{\mp}\nu_e$ ($K_{e3}$) and $K_L\rightarrow e^+e^-e^+e^-$. Results for the last two are preliminary.

The $K_L$ beam was produced every 17~s in 4.8~s spills, $2\times 10^7$ $K_L$s/spill, by 400~GeV protons from CERN SPS.
The $K_L$ production target was located 126~m (more than 20 $c\tau_S$) before the 114~m fiducial volume.

The detector (cf. ref. \cite{na48_cp}) consists of the magnetic and neutral spectrometer.
The magnetic spectrometer consists of 4 drift chambers and bending magnet.
Its momentum resolution is below 1\%. 
Time resolution of the trigger hodoscope is 150~ns.
Neutral decays are measured in the electromagnetic calorimeter hodoscope, filled with liquid Krypton and segmented into 13,212 separately read-out cells.
Its energy resolution is better than 1\%.
The spectrometer also contains a muon veto hodoscope and an additional hadronic calorimeter for independent energy measurement.

\section{$K_L\rightarrow\pi^{\pm}\pi^0 e^{\mp}\nu_e(\bar{\nu_e})$}

Decay matrix element may be expanded in partial waves, parametrized in terms of form factors $F=f_se^{i\delta_s}+f_pe^{i\delta_p}$, $G=ge^{i\delta_p}$ (axial-vector part) and $H=he^{i\delta_p}$ (vector part), and analysed in Cabibbo-Maksymowicz (CM) variables: invariant masses of the dipion $M_{\pi\pi}$ and dilepton $M_{e\nu}$, $\pi^{\pm}$ and $e^{\mp}$ angles $\theta_{\pi}$ and $\theta_e$ relative to the $K_L$ momentum, and angle $\phi$ between the dipion and dilepton planes \cite{cabibbo}.

Two types of triggers were used: the minimum bias trigger for which the requirements for the total energy deposit, hit multiplicity in drift chambers and hits coincidence in the hodoscope were used, and the dedicated $K_{e4}$ trigger where in addition clusters multiplicity in the calorimeter were required.
Efficienies of both triggers were close to 99\%.

The most important cuts defining the $K_{e4}$ sample were those suppressing the $K_L\rightarrow\pi^+\pi^-\pi^0$ ($K_{\pi 3}$) channel ($E/p$ around 1 for electron cadidate and $M_{3\pi}$ {\it vs.} $p_T$ in disagreement with $3\pi$ hypothesis) and $K_{e3}$ with two bremsstrahlung photons ($K_{e3+2\gamma}$) ($M_{2\gamma}$ around $m_{\pi^0}$ and positivity constraint for $\nu$ energy).
Background from these processes was further diminished ($2\%$ for $K_{e3+2\gamma}$ and $1.2\%$ for $K_{3\pi}$) by using neural network algorithm trained on geometric characteristics of showers and tracks. 
Total sample amounted to 5464 $K_{e4}$ with 62 background events.

The branching ratio was determined using normalization to the $K_{\pi 3}$ channel, accounting for trigger efficiencies and Monte Carlo determined acceptances.
It was found $Br(K_{e4})=(5.21\pm 0.07_{stat}\pm 0.09_{syst})\times 10^{-5}$ \cite{na48_ke4}, where the systematic error is dominated by the uncertainty of the $K_{\pi 3}$ branching ratio.
This result is in agreement with previous measurements \cite{carroll,makoff} and has significantly higher accuracy (cf. Fig.~1.).
\begin{figure}
\includegraphics[scale=0.32]{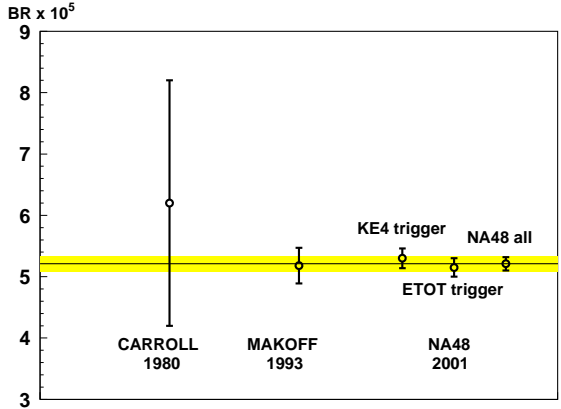}
\vspace{-15mm}
\caption{\em The $K_{e4}$ branching ratio results. The belt refers to the overall NA48 result.}
\vspace{-5mm}
\end{figure} 
The $Br(K_{e4})$ is predominantly sensitive to chiral coupling $L_3$ \cite{amoros}.
Its value was found to be $L_3=(-4.1\pm 0.2)\times 10^{-3}$, in agreement with previous measurement \cite{makoff} and theoretical calculations within chiral perturbation theory \cite{amoros} but with higher accuracy. 

The form factors were determined by fitting differential cross sections to the data in CM variables.
Detector acceptance was accounted for by using Monte Carlo generator with radiative corrections and with flat form factors (cf. Fig. 2).
\begin{figure}[t]
\begin{flushleft}
\begin{tabular}{cc}
\includegraphics[scale=0.22]{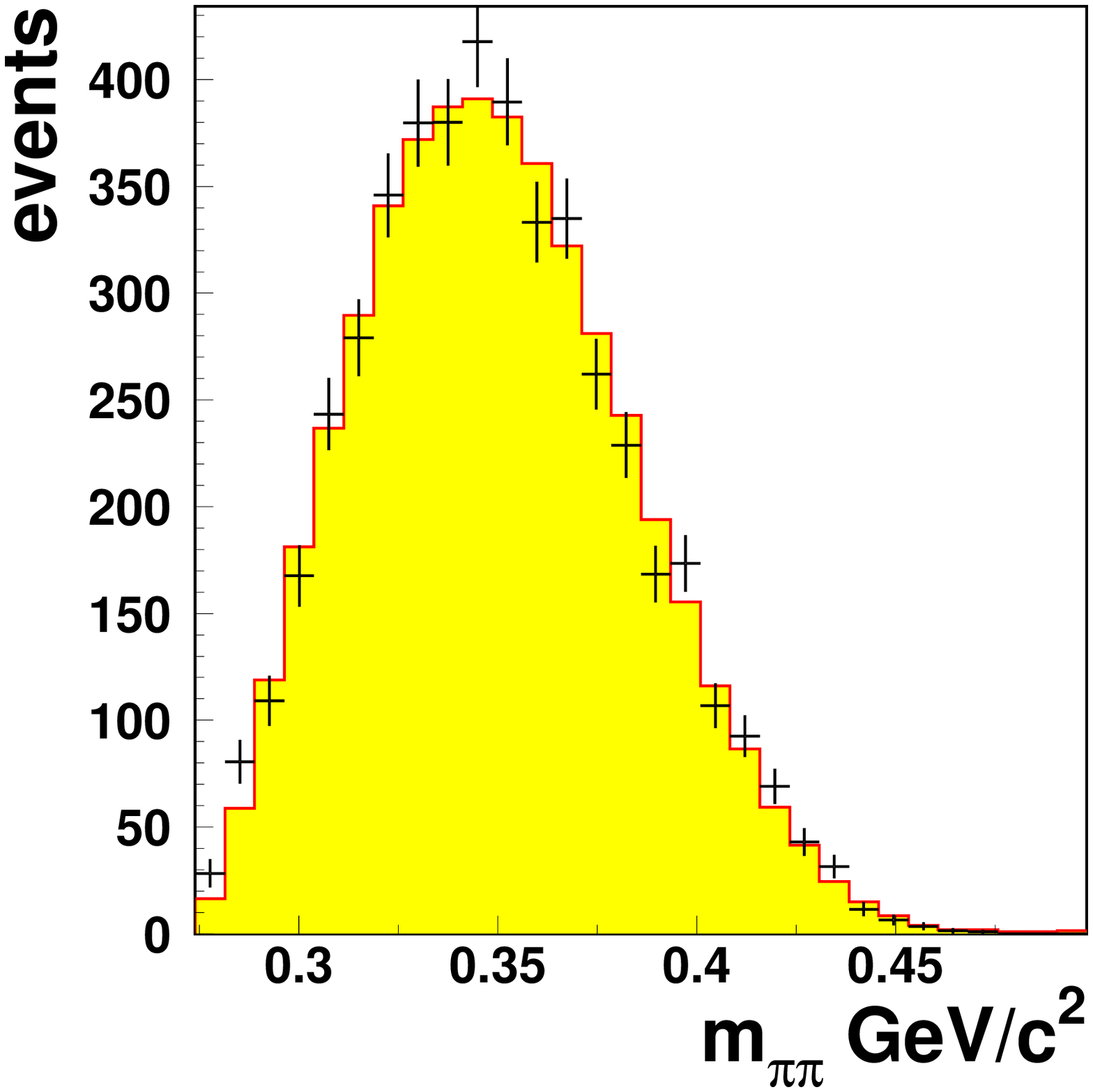} & \includegraphics[scale=0.22]{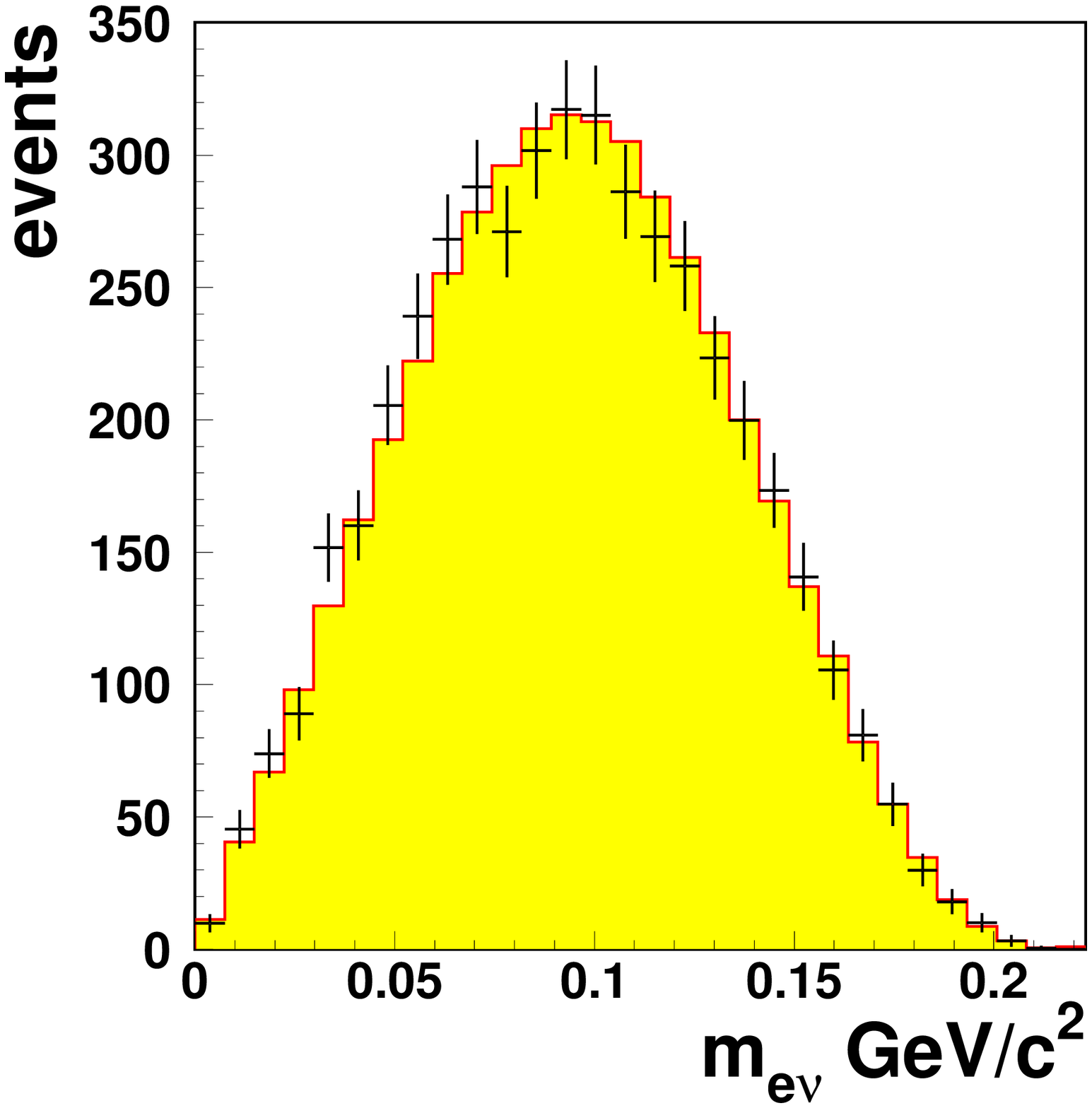} \\
\includegraphics[scale=0.22]{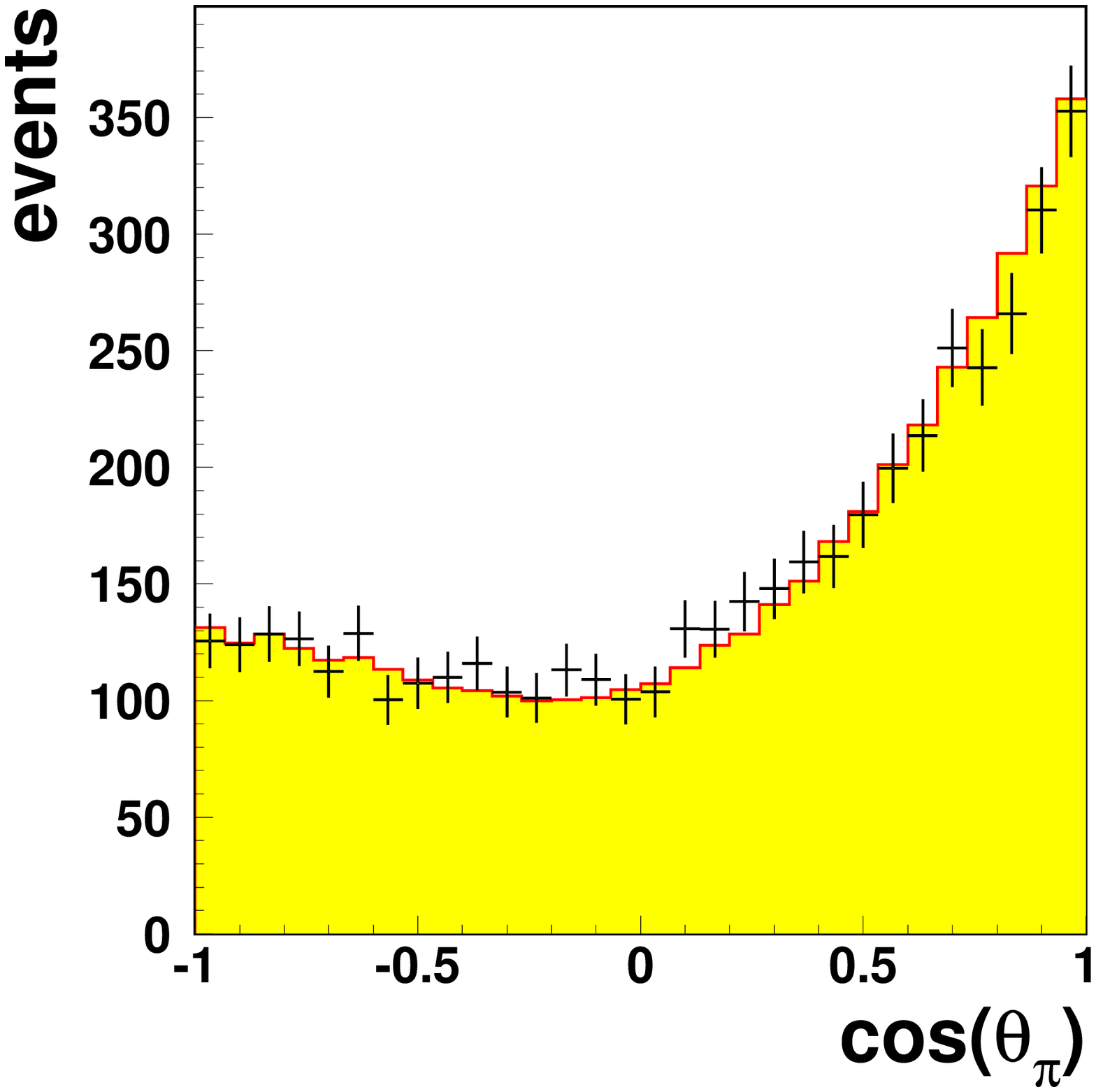} & \includegraphics[scale=0.22]{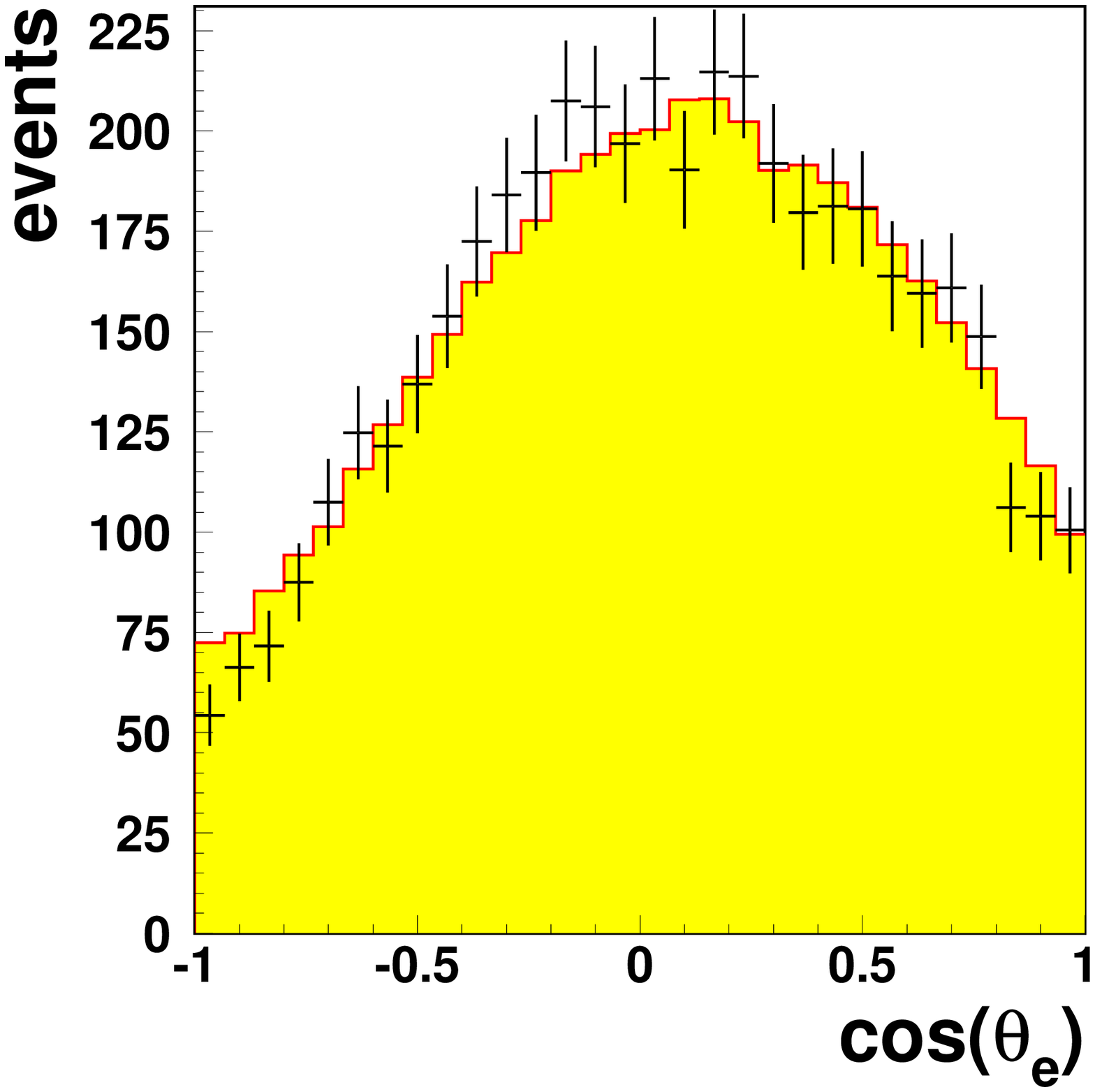} \\ 
\includegraphics[scale=0.22]{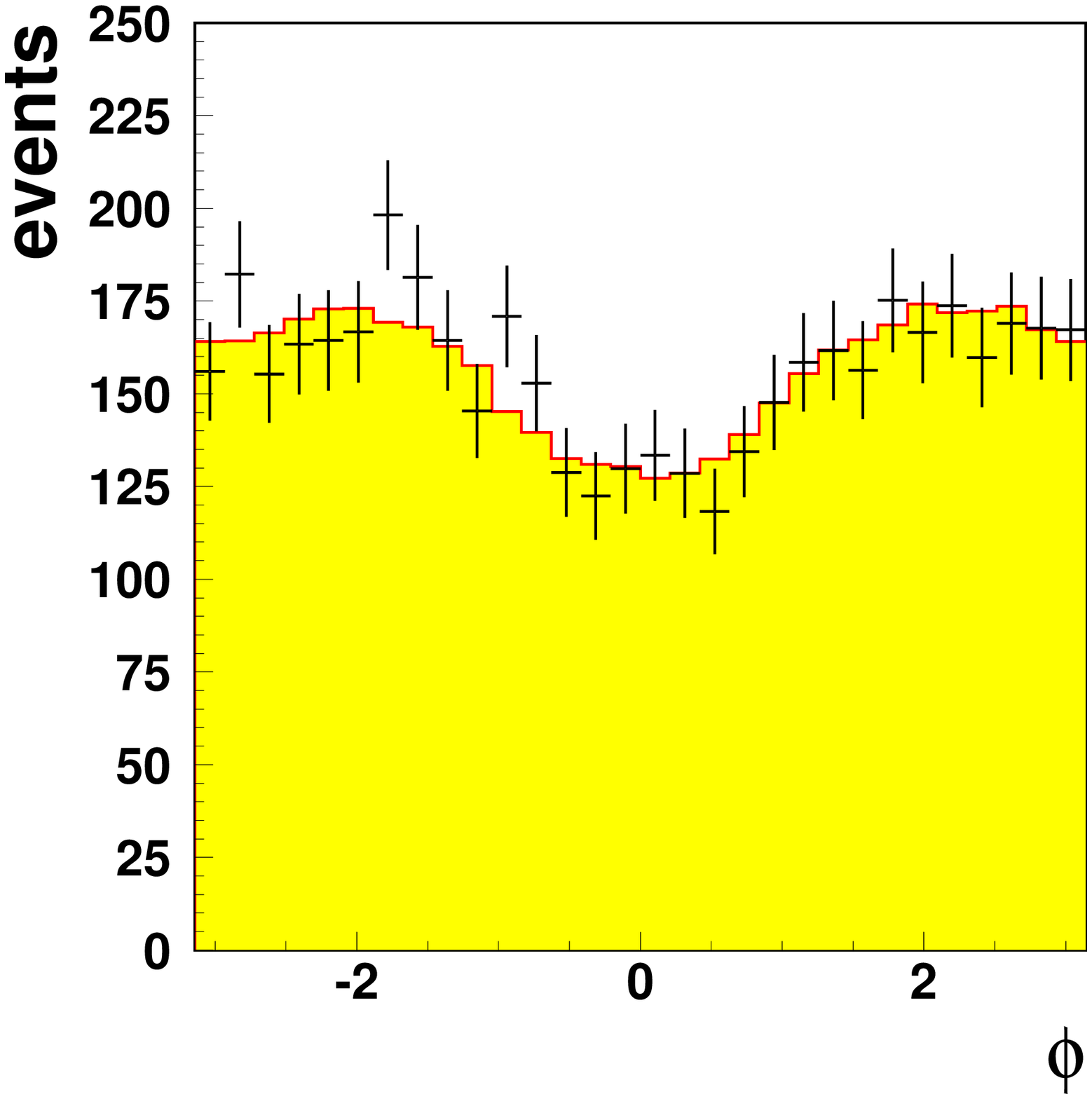} &
\end{tabular}
\vspace{-10mm}
\caption{\em Distributions of the Cabibbo-Maksymowicz variables (data points) with fits (histograms).}
\vspace{-10mm}
\end{flushleft}
\end{figure}
The form factors values obtained from the fit were ($\chi^2/ndf=137/146$)
\begin{eqnarray}
f_s/g & = & 0.052 \pm 0.006_{stat}\pm 0.002_{syst} \nonumber \\
f_p/g & = & -0.051 \pm 0.011_{stat} \pm 0.005_{syst} \nonumber \\
\lambda_g & = & 0.087 \pm 0.019_{stat} \pm 0.006_{syst} \nonumber \\
h/g & = & -0.32 \pm 0.12_{stat} \pm 0.07_{syst}.
\end{eqnarray}
Systematic errors are dominated by the background.
The values are consistent with the FNAL measurements \cite{makoff} and have higher accuracy.
Significantly non-zero value of $f_s/g$ indicates a possiblity of the $\Delta I=1/2$ rule violation at percent level.

\section{$K_L\rightarrow\pi^{\pm}e^{\mp}\nu_e(\bar{\nu_e})$}

Measurement of the $K_{e3}$ decay was motivated by (i) lack of experimental clarity for the non-zero values of non-vector form factors $f_S$ and $f_T$, and (ii) need for a precise test of the unitarity condition $|V_{ud}|^2+|V_{us}|^2+|V_{ub}|^2=1$ by a higher accuracy $|V_{us}|$ mesurement.

Event selection was based on the $E/p>0.93$ requirement for one track. 
The $K_{\pi 3}$ background was reduced to the level $10^{-2}\%$ by cuts on a combined kinematic variables assuming $3\pi$ mass hypothesis.
The muon veto was used for suppressing the $K_{\mu 3}$ background below $10^{-3}\%$.
In total, about 7 millions $K_{e3}$ events were selected.

The form factor analysis was performed by parametrizing the Dalitz distribution as $\rho(E_{\pi}^{\ast},E_e^{\ast})\sim a|V|^2+c|S|^2$, where the vector component was parametrized as $V=f_+(0)(1+\lambda_+q^2/m_{\pi}^2)$ and the non-vector component as $S=f_S+(E_{\nu}^{\ast}-E_e^{\ast})f_T/m_K$, $a$ and $c$ depending on energies and masses only.
Two Dalitz-density analyses were performed: (i) the complete analysis in three variables $E_{\nu}^{\ast}$, $q_1/m_{\pi}^2$ and $q_2/m_{\pi}^2$, and (ii) the two-dimensional analysis using only $q_1/m_{\pi}^2$ and $q_2/m_{\pi}^2$, where pure $V-A$ coupling was assumed.
In the first case (scalar and tensor form factors allowed) the following results were obtained:
\begin{figure}[t]
\begin{flushleft}
\includegraphics[scale=0.4]{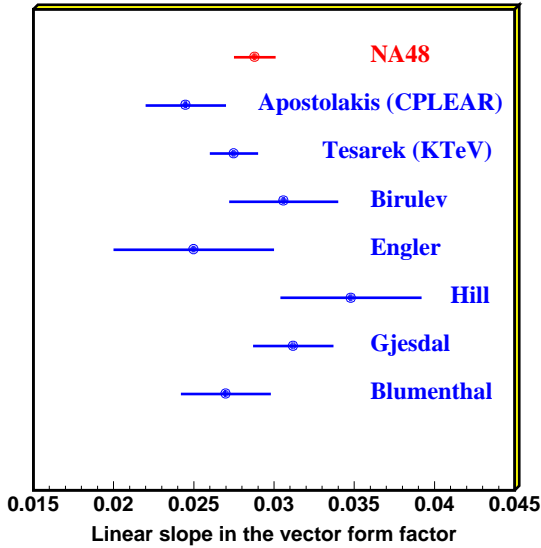}
\vspace{-10mm}
\caption{\em World results for the slope $\lambda_+$ of the vector form factor for the $K_{e3}$ decay.}
\vspace{-5mm}
\end{flushleft}
\end{figure}
\begin{eqnarray}
\lambda_+ & = & 0.0284\pm 0.0007_{stat}\pm 0.0013_{syst} \nonumber \\
|f_S/f_+(0)| & = & 0.015^{+0.007}_{{-0.010}_{stat}}\pm 0.012_{syst} \nonumber \\
|f_T/f_+(0)| & = & 0.05^{+0.03}_{{-0.04}_{stat}}\pm 0.03_{syst} 
\end{eqnarray}
and for pure $V-A$ coupling 
\begin{eqnarray}
\lambda_+ & = & 0.0288\pm 0.0005_{stat}\pm 0.0011_{syst}.
\end{eqnarray}
Both results are at confidence level 70\% and systematic errors are dominanted by the uncertainty of the $K_L$ momentum spectrum.
It is seen that the values of the slope $\lambda_+$ in both analyses are in perfect agreement within errors and the values of $f_S$ and $f_T$ are consistent with zero.
This suggests that there is no significant deviation from the $V-A$ coupling hypothesis, contrary to earlier results from the $K_{e3}^+$ measurement from JINR \cite{akimenko} and with qualitative agreement with many other experimental findings.
Compilation of results for the slope in the vector form factor $\lambda_+$ is presented in Fig. 3.
The NA48 result agrees with the others and exhibits the highest accuracy, exceeding also recent results from other high-rate kaon experiments KTeV \cite{tesarek} and CPLEAR \cite{apostolakis}. 

The $K_{e3}$ branching ratio was also determined using normalization to the $K_L\rightarrow 3\pi^0$ channel.
It was found to be $Br(K_{e3})=0.4010\pm 0.0028_{exp}\pm 0.0035_{norm}$.
Experimental error is dominated by systematic uncertainty of the $K_L$ momentum spectrum. 
Statistical error is negligible.
The normalization error represents the contribution to the systematic error coming from the uncertainty of the $Br(K_L\rightarrow 3\pi^0)$ used for normalization.

It is known that experimental value of the sum $|V_{ud}|^2+|V_{us}|^2+|V_{ub}|^2$ deviates from 1 by $(4.3\pm 1.9)\times 10^{-3}$ and, excluding $|V_{ub}|^2\sim 10^{-5}$, this slight discrepancy may either originate from $|V_{ud}|$ or $|V_{us}|$.
The CKM element $|V_{us}|$ can be determined from the $K_{e3}$ decay rate using relation
\begin{eqnarray}
|V_{us}|f_+^{K^0\pi^-}(0)=\sqrt{\frac{128\pi^3}{G_F^2m_K^5S_{EW}I_K}\Gamma(K_{e3})},
\end{eqnarray}
where the short-distance electroweak factor $S_{EW}=1.0232$ and the phase-space integral $I_K=0.1034\pm 0.0006$.
After correcting for radiative process, by counting radiative events inside Dalitz-plot \cite{cirigliano}, one gets $|V_{us}|f_+^{K^0\pi^-}(0)=0.2139\pm 0.0016$.
Taking average $f_+$ from chiral models: $f_+^{K^0\pi^-}(0)=0.973\pm 0.010$, we obtained $|V_{us}|=0.2199\pm .0016_{exp}\pm 0.0023_{theo}=0.2199\pm 0.0028$.
This is in agreement with the standard model prediction $|V_{us}|=0.2274\pm 0.0021$, assuming the value of $|V_{ud}|=0.9738\pm 0.0005$ \cite{pdg} and the CKM unitarity.
Our result gives us a tighter constraint on $|V_{us}|$ and, together with values for $|V_{ud}|$ and $|V_{ub}|$, persistent discrepancy with the unitarity condition.
Very recently, the KTeV group performed similar studies \cite{ktevus} finding $|V_{us}|=0.2252\pm 0.0022$, consistent with unitarity of the CKM matrix.

\section{$K_L\rightarrow e^+ e^- e^+ e^-$}

This rare decay proceeds via two virtual photons, each converting into $e^+e^-$ pair.
This allows for determination of the form factor for the $K_L\rightarrow\gamma^{\ast}\gamma^{\ast}$ decay vertex, being itself of interest for other $K_L$ leptonic decays.

The 4-track trigger conditions (space points in drift chambers and at least two reconstructed 2-track vertices) were combined with a number of off-line cuts.
Main cuts used for background reduction were those on each track's $E/p>0.9$, reconstructed kaon $p_T^2<5\times 10^{-4}$ GeV$^2$ and $0.475<m_{4e}<0.515$ GeV (cf. Figs. 4 and 5).
\begin{figure}
\includegraphics[width=80mm,height=60mm]{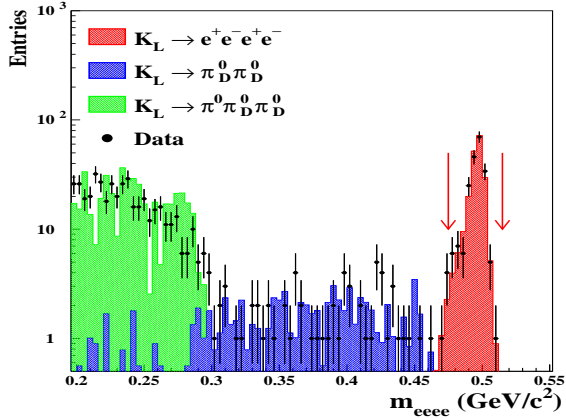}
\vspace{-15mm}
\caption{\em Four-electron mass distributions for Monte Carlo $K_L\rightarrow e^+e^-e^+e^-$ and main background processes $K_L\rightarrow\pi^0_D\pi^0_D$ and $K_L\rightarrow\pi^0\pi^0_D\pi^0_D$ (histograms) and for data (points). Analysis cuts shown with arrows.}
\vspace{-5mm}
\end{figure}
\begin{figure}
\includegraphics[width=80mm,height=60mm]{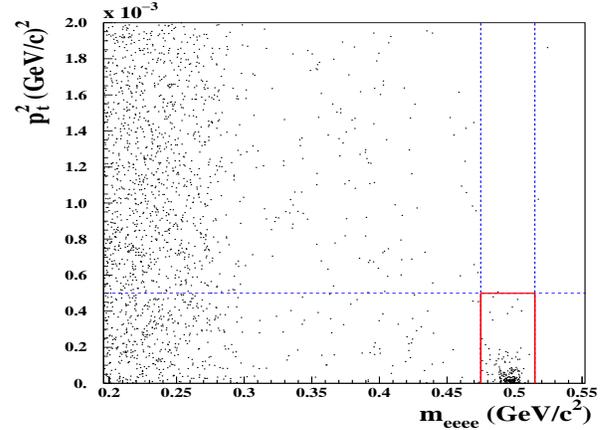}
\vspace{-15mm}
\caption{\em Four-electron $p_T$ vs. mass distribution for data with cuts defining the signal.}
\vspace{-5mm}
\end{figure}
After all selections 200 $K_L\rightarrow e^+ e^- e^+ e^-$ events were found with an estimated background being less than 1\%.
From this, using the $K_L\rightarrow\pi^+\pi^-\pi^0_D$ channel for normalization and Monte Carlo for acceptance corrections, it was found $Br(K_L\rightarrow 4e)=(3.30\pm 0.24_{stat}\pm 0.14_{syst}\pm 0.24_{norm})\times 10^{-8}$.

Earlier measurement performed by KTeV was based on the sample of 441 events with 4.2 background and gave $Br(K_L\rightarrow 4e)=(3.72\pm 0.18_{stat}\pm 0.23_{syst})\times 10^{-8}$ \cite{ktevke4}.
They also reported preliminary result $Br(K_L\rightarrow 4e)=(4.07 \pm 0.12_{stat}\pm 0.11_{syst}\pm 0.16_{norm})\times 10^{-8}$, based on 1056 events \cite{ktev4p}.

\end{document}